\documentclass[prd,groupedaddress,showpacs]{revtex4b5}

\newcommand{\be}{\begin{equation}}
\newcommand{\ee}{\end{equation}}
\newcommand{\ba}{\begin{eqnarray}}
\newcommand{\ea}{\end{eqnarray}}
\newcommand{\bann}{\begin{eqnarray*}}
\newcommand{\eann}{\end{eqnarray*}}
\newcommand{\nn}{\nonumber}
\renewcommand{\l}{\left}
\renewcommand{\r}{\right}

\newcommand{\FThreeTwo}{\text{}_3F_2}
\newcommand{\inttwo}{\int_0^{\Lambda^2}\!\!\! \int_0^\pi  dy \,d\theta\, y  \sin^2\theta}

\usepackage{graphicx}

\begin{document}
\bibliographystyle{apsrev}

\preprint{UNITU-THEP-14/01}

\title{{~\hfill \small \rm UNITU-THEP-14/01\\~\\}
Multiplicative renormalizability of gluon and ghost propagators in QCD}

\author{\vspace{-2ex}J.C.R. Bloch}
\affiliation{Institut f\"ur Theoretische Physik, Universit\"at T\"ubingen, 
Auf der Morgenstelle 14, D-72076 T\"ubingen, Germany\vspace{-1ex}}

\date{\small June 4, 2001}


\begin{abstract}
\vspace{2ex}
We reformulate the coupled set of continuum equations for the renormalized gluon and ghost propagators in QCD, such that the multiplicative renormalizability of the solutions is manifest, independently of the specific form of full vertices and renormalization constants. In the Landau gauge, the equations are free of renormalization constants, and the renormalization point dependence enters only through the renormalized coupling and the renormalized propagator functions. The structure of the equations enables us to devise novel truncations with solutions that are multiplicatively renormalizable and agree with the leading order perturbative results. We show that, for infrared power law behaved propagators, the leading infrared behavior of the gluon equation is not solely determined by the ghost loop, as concluded in previous studies, but that the gluon loop, the three-gluon loop, the four-gluon loop, and even massless quarks  also contribute to the infrared analysis. In our new Landau gauge truncation, the combination of gluon and ghost loop contributions seems to reject infrared power law solutions, but massless quark loops illustrate how additional contributions to the gluon vacuum polarization could reinstate these solutions. Moreover, a schematic study of the three-gluon and four-gluon loops shows that they too need to be considered in more detail before a definite conclusion about the existence of infrared power behaved gluon and ghost propagators can be reached.  
\end{abstract}

\maketitle

\section*{Introduction}

In the Standard Model of the strong, weak and electromagnetic forces, the interactions are quantitatively described by gauge field theories. Quantum chromodynamics is a non-Abelian gauge theory, and the proof of its renormalizability\cite{'tHooft:1971fh} and discovery of ultraviolet asymptotic freedom\cite{Politzer:1973fx} have been milestones in its acceptance as theory of the strong interaction. For large momenta, perturbation theory seems an appropriate calculational tool, as the coupling becomes small. However, for small momenta the coupling grows large and adequate methods have to be used to study nonperturbative phenomena like confinement, chiral symmetry breaking and fermion mass generation\cite{Alkofer:2000wg}. One such method is the study of the Dyson-Schwinger equations\cite{Roberts:1994dr}, and their phenomenological applications to hadronic physics is a subject of growing interest\cite{Roberts:2000aa}. The gluon self-interaction and existence of ghost fields are remarkable features of non-Abelian gauge theories, and the study of the infrared behavior of the gluon and ghost propagators, and of the running coupling has become a major part in the quest to understand color confinement in QCD. 

Early studies of the Dyson-Schwinger equation for the gluon propagator in the Landau gauge concluded that the gluon propagator is highly singular in the infrared\cite{Mandelstam:1979xd,Atkinson:1981er,Atkinson:1982ah,Brown:1989bn}. 
However, these studies neglected any contribution of the ghost fields, and moreover, required one to assume cancellations of certain leading terms in the equations. It is therefore far from certain that these solutions have the correct QCD infrared behavior. That these solutions are at the origin of successful phenomenological applications can be understood by observing that they can generate the necessary integration strength in the kernels of the gap and Bethe-Salpeter equations\cite{Hawes:1998cw}.

Recent studies of the coupled set of continuum Dyson-Schwinger equations for the renormalized gluon and ghost propagators, initiated by von Smekal, Hauck and Alkofer\cite{vonSmekal:1998is}, and later confirmed by Atkinson and Bloch\cite{Atkinson:1998tu,Atkinson:1998zc}, have shown that the coupling of ghost and gluon fields plays a crucial role in the generation of a consistent infrared behavior of QCD. In the infrared the gluon and ghost propagators are power behaved, and the strong running coupling has an infrared fixed point.  The leading infrared exponent is determined by requiring the power solutions to be consistent with both gluon and ghost equations. Although its precise value depends on the vertex Ans\"atze and other details of the truncation, all the results showed that the ghost propagator is more divergent than its bare counterpart, while the gluon propagator is less divergent, and even vanishes as the momentum goes to zero. 
Furthermore, another common feature of these studies is that the leading infrared power in the gluon equation depends only on the ghost loop, and not on the gluon loop. One of our aims is to investigate whether this is a characteristic of QCD or merely an artifact of the approximations.

Multiplicative renormalizability is an important feature of gauge field theories, and it is ruined by the approximations to the renormalized vertices and to the vertex renormalization constants introduced in these studies. Furthermore, the high momentum behavior of the approximate solutions is generally not in agreement with the results of perturbation theory. Herein we aim to develop a truncation scheme, which respects multiplicative renormalizability, and has the correct perturbative limit as well.

Let us first briefly discuss the approximations introduced previously and the defects of their solutions. In Refs.~\cite{Atkinson:1998tu,Atkinson:1998zc} bare vertices are used and the triple-gluon vertex renormalization constant $Z_1$ is set to its bare value of one. This simple approximation allows a straightforward basic analysis of the coupled set of ghost and gluon equations. As has been shown in detail in Ref.~\cite{Atkinson:1998tu}, this approximation violates the multiplicative renormalizability of the solutions, and furthermore, neither the anomalous dimension of their logarithmic ultraviolet behavior nor the leading order beta coefficient of the running coupling  satisfy the results of perturbation theory. 
The authors were, however, not able to conclude if this slight, but apparent, discrepancy with perturbation theory was due to an artifact of their approximation or if it could be a genuine feature of full QCD. In Ref.~\cite{vonSmekal:1998is} more care was given to the construction of the vertices, requiring the Ward-Takahashi identities to be satisfied to some extent. 
However, to force the solutions to have the correct perturbative behavior, the authors introduced a manipulation of the renormalization constant $Z_1$, which again violates the multiplicative renormalizability of the solutions.

In Sections (III-VIII) of Ref.~\cite{Atkinson:1998tu}, an intriguing truncation was studied: based on symmetry ideas between the ghost and gluon equations only the ghost loop is retained in the gluon equation. The paper showed in detail that the solutions of this truncated set of equations are multiplicatively renormalizable and that their high momentum behavior satisfies the perturbative results of this truncated theory. Although this truncation is not consistent with QCD, as the gluon loop was discarded, its solutions exhibit these important properties expected from solutions of gauge field theories, thereby providing insight into how multiplicative renormalizability of the nonperturbative solutions is achieved, how the high momentum behavior of these solutions obey the leading order perturbative results, and even how both features are entangled. Section (IX) of Ref.~\cite{Atkinson:1998tu} then went on to show how a naive treatment of the gluon loop destroys both multiplicative renormalizability and agreement with perturbation theory.

In addition to the above-mentioned approximations, Refs.~\cite{vonSmekal:1998is,Atkinson:1998tu} also introduce angular approximations to make the analytical and numerical calculations more tractable.  These angular approximations are satisfying for slowly varying functions\cite{Landau:1956}, like propagator functions in their perturbative regime. However, Ref.~\cite{Atkinson:1998zc} showed that these angular approximations can introduce a significant error for regions where the functions are power behaved. Although this means that eventually the analysis of the asymptotic infrared power behavior will require the angular integrations to be performed exactly, the qualitative conclusions about multiplicative renormalizability and high momentum behavior of the solutions are, however, not affected by the angular approximation. 

In this paper we reformulate the coupled set of continuum equations for the renormalized gluon and ghost propagators, such that the Landau gauge equations are free of renormalization constants, and depend on the renormalization point only through the renormalized coupling and the renormalized propagator functions. In analogy to the ghost-loop only truncation of Ref.~\cite{Atkinson:1998tu}, we see how the simple structure of the equations makes the multiplicative renormalizability transparent, i.e. not hidden in renormalization constants or renormalized vertices.  We then introduce a novel truncation scheme to these formal equations, which respects both the multiplicative renormalizability and the perturbative limit of the solutions. We show that, for power law solutions, the leading infrared behavior of the gluon equation no longer solely depends on the ghost loop, but rather depends on the ghost loop, the gluon loop, the three-gluon loop, the four-gluon loop and even on massless quark loops. The consequences for the existence of infrared power behaved propagators are investigated in the Landau gauge. We also briefly develop the same ideas for the massless quark equation.

\section*{The equations}

The QCD Dyson-Schwinger equations for the gluon and ghost propagators are (provisionally neglecting the four-gluon vertex and setting the number of quarks to zero):
\ba\label{SD-Gl}
\l[D_{\mu\nu}(p)\r]^{-1} = \l[D^0_{\mu\nu}(p)\r]^{-1} 
&-& (-1) N_c g_0^2 \int \frac{d^4 q}{(2\pi)^4} \, G^0_\mu(-r, q) \Delta(q) G_\nu(q, -r) \Delta(-r) \\
&-& \frac{1}{2}N_c g_0^2 \int \frac{d^4 q}{(2\pi)^4} \, \Gamma^{3g,0}_{\mu\alpha_1\alpha_2}(-p,q,r) D^{\alpha_1\beta_1}(q) \Gamma^{3g}_{\nu\beta_2\beta_1}(p,-r,-q) D^{\alpha_2\beta_2}(r) \,,\nn \\
\label{SD-Gh}
\l[\Delta(p)\r]^{-1} = \l[\Delta^0(p)\r]^{-1} &-& N_c g_0^2 \int \frac{d^4 q}{(2\pi)^4} \, G^0_\mu(p,q) \Delta(q) G_\nu(q,p) D^{\mu\nu}(r)  \,,
\ea
where $g_0$ is the bare coupling, $D_{\mu\nu}$ the gluon propagator, $\Delta$ the ghost propagator, $\Gamma_{\alpha\beta\gamma}^{3g}$ the triple-gluon vertex, $G_\mu$ the ghost-gluon vertex, the zero superscript denotes bare quantities, and $r=p-q$. Note that the contraction over color indices has already been performed, and the extra factor $(-1)$ in front of the ghost loop is due to the Fermi statistics of the ghost field.

The general expressions for the full gluon and ghost propagators in a covariant gauge $\xi$ can be written as
\ba\label{Full-Gl}
D_{\mu\nu}(p) &=&  -i \left[ \left(g_{\mu\nu}-\frac{p_\mu p_\nu}{p^2}\right) \frac{F(p^2)}{p^2} + \xi \frac{p_\mu p_\nu}{p^4} \right]\,,\\
\label{Full-Gh}
\Delta(p) &=& \frac{i \, G(p^2)}{p^2} \,,
\ea
and we shall refer to the scalar functions $F(p^2)$ and $G(p^2)$ as the gluon and ghost form factors.

Although we do not know the correct expressions for the full triple-gluon and gluon-ghost vertices, we can always formally rewrite Eqs.~(\ref{SD-Gl}, \ref{SD-Gh}), for arbitrary vertices, as:
\begin{eqnarray}\label{SD3}
\frac{1}{F(x)} &=& 1 - \frac{N_c g_0^2}{8 \pi^3}
\inttwo \, \l[ M(x,y,z) \, G(y) G(z) +  Q(x,y,z) \, F(y) F(z)\r]
 \,,\\
\label{SD4}
\frac{1}{G(x)} &=& 1
- \frac{N_c g_0^2}{8 \pi^3} \inttwo \, T(x,y,z) \, G(y) F(z) \,.
\end{eqnarray}
where we first contracted Eq.~(\ref{SD-Gl}) with $-i{\cal P}^{\mu\nu}(p)/3p^2$, where ${\cal P}^{\mu\nu}(p) = g^{\mu\nu}-4 p^\mu p^\nu/p^2$, thus avoiding spurious quadratic ultraviolet divergences\cite{Brown:1989bn}, and multiplied Eq.~(\ref{SD-Gh}) with $i/p^2$. Subsequently we applied a Wick rotation to both equations, which introduces an imaginary factor $i$ for each four dimensional momentum integration, evaluated two trivial angular integrations, and set $x=p^2$, $y=q^2$, $z=r^2$. 
The kernels $M, Q, T$ depend on the full vertices, and are defined by:
\ba
\label{kernelM} M(p^2,q^2,r^2) &=&  \frac{1}{3 p^2 \,q^2\, r^2} \,  {\cal P}^{\mu\nu}(p) \,G^0_\mu(-r, q) \, G_\nu(q, -r) \,,\\
\label{kernelQ} Q(p^2,q^2,r^2) &=& - \frac{1}{6 p^2 q^2 r^2} \, {\cal P}^{\mu\nu}(p) \,  \Gamma^{3g,0}_{\mu\alpha_1\alpha_2}(-p,q,r) \,\Gamma^{3g}_{\nu\beta_2\beta_1}(p,-r,-q) \, \\ && \times \left[ g_\perp^{\alpha_1\beta_1}(q) \, g_\perp^{\alpha_2\beta_2}(r) + \xi \left( g_\perp^{\alpha_1\beta_1}(q)\frac{r^{\alpha_2} r^{\beta_2}}{r^2 F(r^2)} + g_\perp^{\alpha_2\beta_2}(r)  \frac{q^{\alpha_1} q^{\beta_1}}{q^2 F(q^2)} \right) + \xi^2 \frac{q^{\alpha_1} q^{\beta_1}}{q^2 F(q^2)} \frac{r^{\alpha_2} r^{\beta_2}}{r^2 F(r^2)}\right] \,,\nn\\
\label{kernelT} T(p^2,q^2,r^2) &=& - \frac{1}{p^2 \, q^2 \, r^2} \, \left[g_\perp^{\mu\nu}(r) + \xi \frac{r^\mu r^\nu}{r^2 F(r^2)} \right] \, G^0_\mu(p,q) \, G_\nu(q,p) \,,
\ea
where $g^{\mu\nu}_\perp(q) = g^{\mu\nu} - q^\mu q^\nu/q^2$. Note that the full vertices will depend on the propagator functions $F$ and $G$ by the intermediary of the three-point Dyson-Schwinger equations.

The full, regularized, unrenormalized Green's functions are potentially divergent as we take the ultraviolet cutoff $\Lambda$ to infinity, and we therefore introduce renormalized gluon and ghost fields, and a renormalized coupling. The renormalized gluon and ghost form factors $F_R$ and $G_R$ satisfy the following multiplicative renormalization condition:
\be
F(p^2, \Lambda^2) = Z_3(\mu^2,\Lambda^2) F_R(p^2, \mu^2) \,, \hspace{1cm}
G(p^2, \Lambda^2) = \tilde Z_3(\mu^2,\Lambda^2) G_R(p^2, \mu^2)  \,,
\label{renorm}
\ee
where $F_R(\mu^2, \mu^2)=G_R(\mu^2, \mu^2) \equiv 1$ and $Z_3$, $\tilde Z_3$ are the renormalization constants for the gluon and ghost fields. Because of gauge invariance the renormalized coupling $g$ satisfies
\be
g(\mu^2) = \frac{Z_3^{3/2}(\mu^2,\Lambda^2)}{Z_1(\mu^2,\Lambda^2)} \, g_0(\Lambda^2) = \frac{Z_3^{1/2}(\mu^2,\Lambda^2)\tilde
Z_3(\mu^2,\Lambda^2)}{\tilde{Z}_1(\mu^2,\Lambda^2)} \, g_0(\Lambda^2) \,,
\label{renormg}
\ee
where $Z_1$, $\tilde Z_1$ are the renormalization constants for the triple-gluon and gluon-ghost vertices. From Eqs.~(\ref{renorm}, \ref{renormg}) it is easy to see\cite{Atkinson:1998tu} that the product
\be
\label{rgi2}
\hat\alpha(x,\Lambda^2) \equiv  \frac{g^2(\mu^2)}{4\pi} \tilde Z_1^2(\mu^2,\Lambda^2) F_R(x,\mu^2) G_R^2(x,\mu^2) = \hat\alpha(\mu^2,\Lambda^2) F_R(x,\mu^2) G_R^2(x,\mu^2) \,
\ee
is renormalization group invariant, i.e. independent of $\mu$.

We know from Taylor\cite{Taylor:1971ff} that  $\tilde Z_1(\mu^2,\Lambda^2) = 1$ in the Landau gauge, and there the renormalization group invariant quantity $\hat\alpha(x,\Lambda^2)$ of Eq.~(\ref{rgi2}) is independent of $\Lambda^2$, and nothing else but the running coupling $\alpha(x)= g^2(x)/4\pi$. Although all our formal derivations will be valid for any covariant gauge, in the following we will focus on the Landau gauge and systematically replace $\hat\alpha(x,\Lambda^2)$ by $\alpha(x)$, keeping in mind that $\hat\alpha(x,\Lambda^2)$ is the correct renormalization group invariant quantity to be used when studying the equations in an arbitrary covariant gauge. 

After introducing renormalized quantities in Eq.~(\ref{SD4}) using Eqs.~(\ref{renorm}, \ref{renormg}, \ref{rgi2}),  we find the following equation for the renormalized ghost form factor:
\be
\label{SD6}
\frac{1}{G_R(x)} = \tilde Z_3(\mu^2,\Lambda^2)
- \frac{N_c \,\alpha_\mu}{2 \pi^2} \inttwo \, T(x,y,z)\, G_R(y) F_R(z) \,,
\ee
where $\alpha_\mu$ denotes the value $\alpha(\mu^2)$ of the coupling at the renormalization point.

In an analogous way we derive an equation for the renormalized gluon form factor from Eq.~(\ref{SD3}):
\begin{eqnarray}\label{SD5}
\frac{1}{F_R(x)} &=& Z_3(\mu^2,\Lambda^2) - \frac{N_c g^2(\mu^2)}{8 \pi^3} 
\inttwo \, \\
 && \times \l[ \tilde Z_1^2(\mu^2,\Lambda^2) M(x,y,z) G_R(y) G_R(z) + Z_1^2(\mu^2,\Lambda^2) Q(x,y,z) F_R(y) F_R(z) \r] \nn\,.
\end{eqnarray}

From Eq.~(\ref{renormg}) we see that:
\begin{equation}
Z_1(\mu^2,\Lambda^2) = \frac{Z_3(\mu^2,\Lambda^2)}{\tilde Z_3(\mu^2,\Lambda^2)} \tilde Z_1(\mu^2,\Lambda^2) \,,
\end{equation}
and after using Eq.~(\ref{renorm}) we can write this as:
\begin{equation}\label{Z1ratio}
Z_1(\mu^2,\Lambda^2) = \frac{F(v,\Lambda^2)}{F_R(v,\mu^2)}\frac{G_R(w,\mu^2)}{G(w,\Lambda^2)} \tilde Z_1(\mu^2,\Lambda^2) , \hspace{.5cm} \text{for arbitrary } v,w \le \Lambda.
\end{equation}
This expression enables us to factorize $Z_1$ in a $\mu$ and a $\Lambda$ dependent part in the Landau gauge ($\tilde Z_1 = 1$). Although $Z_1$ depends only on $\mu$ and $\Lambda$, we can always write it as the ratio (\ref{Z1ratio}) with arbitrary momenta $v$ and $w$. We eliminate the factor $Z_1^2$ in Eq.~(\ref{SD5}) by substituting Eq.~(\ref{Z1ratio}) twice, once choosing $v=w\equiv y$, and once with $v=w\equiv z$. Hence,
\begin{equation}\label{SD7}
\frac{1}{F_R(x)} = Z_3(\mu^2,\Lambda^2) - \frac{N_c \alpha_\mu}{2 \pi^2} 
\inttwo \, R(x,y,z) \,  G_R(y) G_R(z)\,,
\end{equation}
where we have introduced the notation:
\begin{equation}\label{Rxyz}
R(x,y,z) =  M(x,y,z) + \frac{F(y,\Lambda^2)F(z,\Lambda^2)}{G(y,\Lambda^2)G(z,\Lambda^2)} \, Q(x,y,z) \,.
\end{equation}
When introducing the renormalized coupling $g(\mu^2)$ in the gluon loop of Eq.~(\ref{SD5}) we found it most logical to use the $Z_1$-identity of Eq.~(\ref{renormg}). An equivalent, more direct, way is to use the $\tilde Z_1$-identity instead. Obviously the same equation, Eq.~(\ref{SD7}), results.

The two remaining renormalization constants $Z_3$ and $\tilde Z_3$ in Eqs.~(\ref{SD6}, \ref{SD7}) can now easily be eliminated by subtracting each of the equations at two different momenta: 
\ba\label{SD9}
\frac{1}{F_R(x)} &=& \frac{1}{F_R(\sigma)} - \frac{N_c \alpha_\mu}{2 \pi^2} 
\inttwo \, \left[ \bigg( R(x,y,z) \,  G_R(y) G_R(z)\bigg) -  \bigg(x \leftrightarrow \sigma \bigg) \right]\,, \\
\label{SD10}
\frac{1}{G_R(x)} &=& \frac{1}{G_R(\sigma)}
- \frac{N_c \,\alpha_\mu}{2 \pi^2} 
\inttwo \, \left[ \bigg( T(x,y,z)\, G_R(y) F_R(z)\bigg) - \bigg(x \leftrightarrow \sigma\bigg) \right] \,,
\ea
where the momentum $\sigma$ can, but does not have to be taken to coincide with the renormalization point $\mu$.
Let us now have a closer look at Eqs.~(\ref{SD9}, \ref{SD10}). First of all we observe that, in the Landau gauge, the equations are free of renormalization constants, and that the only renormalization point dependence comes through the renormalized coupling $\alpha_\mu$ and the renormalized propagator functions $F_R(\ast,\mu^2)$ and $G_R(\ast,\mu^2)$. Indeed, as can be seen from Eqs.~(\ref{kernelM}, \ref{kernelQ}, \ref{kernelT}, \ref{Rxyz}), the unspecified kernels $R(x,y,z)$ and $T(x,y,z)$ are to be calculated using the triple-gluon and ghost-gluon vertices, and the gluon and ghost form factors, all in their full, regularized, unrenormalized form, and are therefore independent of $\mu$. 

On the other hand, we also know that, except for the upper integration limit, the kernels $R$ and $T$ are the only $\Lambda$-dependent factors remaining in the equations in the Landau gauge. If multiplicative renormalizability is to be a general nonperturbative feature of gauge theories, then $F_R(x)$ and $G_R(x)$ will remain finite as $\Lambda$ is taken to infinity. Therefore the kernels of the integrals in Eqs.~(\ref{SD9}, \ref{SD10}) are such that after subtraction and integration the $\Lambda$-dependence vanishes as $\Lambda$ is taken to infinity. Moreover, the $\Lambda$-dependences of the integrals in the unsubtracted equations, Eqs.~(\ref{SD6}, \ref{SD7}), correspond to those of $\tilde Z_3$ and $Z_3$ known from perturbation theory. This can easily be verified from the $\Lambda$-dependence of the various elements in the equations given by the renormalization group equations. Any relevant approximation to the propagator equations will obviously have to respect this property.  

 Let us show how the multiplicative renormalizability of the solutions to Eqs.~(\ref{SD9}, \ref{SD10}) is satisfied.  From Eq.~(\ref{renorm}) we see that for solutions to be multiplicatively renormalizable they have to satisfy 
\be
\label{munu}
F_R(x, \nu^2) = \frac{F_R(x,\mu^2)}{F_R(\nu^2, \mu^2)} \,, \hspace{15mm}
G_R(x, \nu^2) = \frac{G_R(x,\mu^2)}{G_R(\nu^2, \mu^2)} \,,
\ee
when renormalized at different scales $\mu^2$ and $\nu^2$. It is straightforward to see,  using Eq.~(\ref{rgi2}), that \textit{if} $F_R(x,\mu^2)$ and $G_R(x,\mu^2)$ are solutions of the set of equations (\ref{SD9}, \ref{SD10}) renormalized at $\mu^2$, \textit{then} indeed $F_R(x,\nu^2)$ and $G_R(x,\nu^2)$ defined by Eq.~(\ref{munu}) will satisfy the same equations now renormalized at $\nu^2$. It is precisely using this reasoning that one can show that the approximations introduced in Refs.~\cite{vonSmekal:1998is,Atkinson:1998tu,Atkinson:1998zc} to treat the gluon loop are violating multiplicative renormalizability. Note that \textit{any} momentum scale can be chosen as renormalization point; it is not restricted to the small coupling region of perturbation theory.

Of course, the new gluon equation, Eq.~(\ref{SD7}), is completely equivalent to the original one,  Eq.~(\ref{SD5}), both having the same multiplicatively renormalizable solutions. So what are the advantages of the new formulation? In contrast to previous studies, we have eliminated the renormalization constant $Z_1$ without introducing any approximation yet, and therefore multiplicative renormalizability is preserved. We will show how this allows us to devise tractable approximations to the equations, having solutions that will still respect multiplicative renormalizability.  With the new approximation we will then try to answer the following questions. Do the new nonperturbative solutions agree with perturbation theory for high momenta?  Do the solutions behave as power laws in the infrared, and is it indeed so that only the ghost loop contributes to the leading infrared power in the gluon equation as deduced in Refs.~\cite{vonSmekal:1998is,Atkinson:1998tu,Atkinson:1998zc}? In fact, from Eqs.~(\ref{SD9}, \ref{SD10}) one can introduce novel truncations to the equations for the renormalized gluon and ghost propagator that will readily contradict this statement. 
Although we know that the bare vertex approximation from Ref.~\cite{Atkinson:1998tu} does not preserve multiplicative renormalizability, we can now define an alternative bare truncation using Eqs.~(\ref{SD9}, \ref{SD10}) that will preserve both multiplicative renormalizability and the correct perturbative limit by simply taking the bare approximation to the complete kernels $R$ and $T$. The bare approximation is defined by substituting
\be\label{bare-elems}
\Gamma_{\alpha\beta\gamma}^{3g} = \Gamma_{\alpha\beta\gamma}^{3g,0} \,, \qquad G_\mu = G_\mu^0 \,, \qquad F = F_0 = 1 \,, \qquad G = G_0 = 1
\ee
in Eqs.~(\ref{kernelM}, \ref{kernelQ}, \ref{kernelT}, \ref{Rxyz}), hence yielding
\be\label{bare-approx}
R(x,y,z) \approx R_0(x,y,z) = M_0(x,y,z) + Q_0(x,y,z) \,, \hspace{1cm}
T(x,y,z) \approx T_0(x,y,z) \,.
\ee
The expressions for $R_0$ and $T_0$ are given in Eqs.~(\ref{Rxyz0}, \ref{Txyz0}). For the gluon loop the approximation amounts to assume that the ratio of unrenormalized form factors $F(y)F(z)/G(y)G(z)$ in Eq.~(\ref{Rxyz}) cancels the full unrenormalized corrections to the bare triple-gluon vertex to a good approximation, at least after contraction and integration. In fact this could be considered as an extension to the approximation introduced by Mandelstam\cite{Mandelstam:1979xd}. There the Ward identity was used to simplify the product of full vertex and full propagators. Our study now also includes the renormalization aspects of the gluon loop, which were omitted before. We propose to  analyze  the validity of our assumption in a study of the 3-point Dyson-Schwinger equation in future work. 

In Section (IX) of Ref.~\cite{Atkinson:1998tu} it was shown that consistency with perturbation theory in the high momentum region is not trivially satisfied in a naive treatment of the gluon loop: even if the truncation leaves invariant the perturbative expansion of the equations, the high momentum limit of the nonperturbative solutions usually does not agree with the results of perturbation theory. However, the same study pointed out that, in the bare vertex "ghost-loop only" truncation, the structure of the loop integrands ensuring multiplicative renormalizability is exactly what is needed such that the high momentum limit of the nonperturbative solutions be consistent with the results of perturbation theory, i.e. it yields the correct values for the anomalous dimensions of the gluon and ghost propagators, and for the leading order beta coefficient of the running coupling. All these arguments now also apply to the bare approximation to Eqs.~(\ref{SD9}, \ref{SD10}), which now include the gluon loop, and the nonperturbative solutions do agree with the leading order perturbative results in the high momentum region, as  is illustrated in more detail in Appendix~\ref{App:bare}.
The nonperturbative solutions have a leading ultraviolet logarithmic behavior 
\be\label{uv}
F_R(x) \sim \log^\gamma x\,, \hspace{1cm}
G_R(x) \sim \log^\delta x \,, \hspace{1cm}
\alpha(x) \sim \frac{4\pi}{\beta_0\log x/\Lambda_{QCD}^2} \,,
\ee
with $\beta_0 = 11N_c/3$,  $\gamma=-13N_c/6\beta_0=-13/22$, and $\delta=-3N_c/4\beta_0=-9/44$,
thus reproducing the correct perturbative results for the anomalous dimensions of the gluon and ghost propagators, and for the leading order beta coefficient of the running coupling. Note that by satisfying the perturbative limit, the approximation also ensures that the unsubtracted equations Eqs.~(\ref{SD6}, \ref{SD7}) yield the correct $\Lambda$-dependence for $\tilde Z_3$ and $Z_3$.

 Next we analyze the infrared behavior of Eqs.~(\ref{SD9}, \ref{SD10}) in our bare approximation. Following the arguments of Ref.~\cite{Atkinson:1998zc} we deduce that, as before, both gluon and ghost equations are individually satisfied by propagators which behave as power laws in the infrared, 
\be\label{powlaw}
F_R(x) \sim x^{2\kappa} \,, \hspace{1cm} G_R(x) \sim x^{-\kappa} \,, \hspace{1cm} \alpha(x) \to \text{constant} \,,
\ee
and, because of Eq.~(\ref{rgi2}), these power laws lead to an infrared fixed point for the running coupling,
\be\label{nu}
\alpha_0 = \lim_{x \to 0} \alpha_\mu F_R(x) G_R^2(x) \,.
\ee

 After substitution of the power laws (\ref{powlaw}) in Eqs.~(\ref{SD9}, \ref{SD10}), the ghost and gluon equations each yield a relation between the infrared fixed point $\alpha_0$ and the leading infrared exponent $\kappa$:  \be\label{condit}
N_c \, \alpha_0  = \frac{1}{\chi_{gh}(\kappa)} \,, \qquad \qquad N_c \, \alpha_0 = \frac{1}{\chi_{gl}(\kappa)} \,,
\ee
 which are derived by equating the coefficients of the leading power of $x$ for $x \to 0$ on both sides of each equation, and where $\chi_{gh}(\kappa)$, $\chi_{gl}(\kappa)$ are computed by solving the loop integrals as detailed in Ref.~\cite{Atkinson:1998zc}. A consistent infrared power solution requires the gluon and ghost equations to be satisfied simultaneously,
\be\label{consist}
\chi_{gh}(\kappa) = \chi_{gl}(\kappa) \,.
\ee
The solution of this equation yields the value of the leading infrared exponent $\kappa$, and the corresponding $\alpha_0$ can then be computed from Eq.~(\ref{condit}).

It is important to note that, in contrast to the studies \cite{vonSmekal:1998is,Atkinson:1998tu,Atkinson:1998zc}, the gluon loop now also contributes to the leading order infrared power of the gluon equation, Eq.~(\ref{SD9}), and hence to $\chi_{gl}(\kappa)$, in our novel truncation.
The expressions for $\chi_{gh}(\kappa)$, $\chi_{gl}(\kappa)$ resulting from the leading infrared analysis in the bare approximation are expressed in terms of generalized hypergeometric functions, and are given in Eqs.~(\ref{nugh}, \ref{nugl}). Numerical evaluation, using Mathematica, has shown that $\chi_{gl}(\kappa) < 0$ for any $\kappa$ for which the integrals converge. This disagrees with the ghost equation for which $\chi_{gh}(\kappa) > 0$, and hence no consistent solution for $\kappa$ exists, and power laws are not infrared asymptotic solutions of Eqs.~(\ref{SD9}, \ref{SD10}) in the bare approximation. Note that in previous studies\cite{vonSmekal:1998is,Atkinson:1998tu,Atkinson:1998zc}, in which the gluon loop was subleading and only the ghost loop contributed to the gluon vacuum polarization, the power law solutions were consistent and the value of $\kappa$ varied between 0.7 and 0.9 depending on the truncation. In our study the gluon equation seems to reject the power law solutions, and although this could at first be seen as a setback, we do not believe that this invalidates the general ideas of our approach. We rather believe it is an artifact of the truncation, and more care has to be taken in the analysis of the various contributions to the gluon vacuum polarization, as we will show below. A case can be made for the eventual survival of infrared power laws by observing that $\chi_{gh}(\kappa)$, determined from the ghost equation, can be identified with the reciprocal of the infrared fixed point of the running coupling (using Eq.~(\ref{condit})), and that its numerical values can be considered physical for $\kappa$ in the convergence region of the integrals, e.g. it varies from 0.87 to 0.24 for $\kappa$ varying from 0.2 to 1.

Several contributions to the gluon vacuum polarization were neglected in the gluon equation, Eq.~(\ref{SD-Gl}),  and we now extend the scope of our investigation by considering its complete form:
\be\label{SD-Gl-full}
\l[D_{\mu\nu}(p)\r]^{-1} = \l[D^0_{\mu\nu}(p)\r]^{-1} 
- \pi_{\mu\nu}^{gh}(p) - \pi_{\mu\nu}^{g}(p) - \pi_{\mu\nu}^{3g}(p) - \pi_{\mu\nu}^{4g}(p) - \pi_{\mu\nu}^{tad}(p) - \pi_{\mu\nu}^{q}(p)  \,,
\ee
including the contributions from the ghost loop, gluon loop, three-gluon loop, four-gluon loop, tadpole diagram and quark loop. First, we will show that massless quarks can contribute to the leading infrared power analysis, and how they can affect the existence of a consistent infrared behavior. Although quark masses will most probably alter these conclusions, the main purpose of this exercise is to illustrate, in a simple way, that the absence of a consistent leading infrared power $\kappa$, resulting from the multiplicatively renormalizable treatment  of the gluon loop, does not have to be definitive; other contributions, overlooked till now, may reverse this situation. In fact, the hope is that this will be realized by the three-gluon and four-gluon loops, both involving the four-gluon vertex. Although no numerical calculation could yet be performed, we will schematically show that these loops can also contribute to leading order in the infrared, even though they are only subleading in the ultraviolet. 

Consider the quark loop contribution to the gluon vacuum polarization in Eq.~(\ref{SD-Gl-full}):
\be\label{piq}
\pi^q_{\mu\nu}(p) = (-1) \frac{N_f}{2} g_0^2 \int \frac{d^4 q}{(2\pi)^4} \, \text{Tr} \l[ \Gamma^{qg,0}_{\mu}(-r,q,-p) S_F(q) \Gamma^{qg}_{\nu}(q,-r,p) S_F(-r) \r] \,,
\ee
where $S_F$ is the quark propagator, $\Gamma^{qg}_{\mu}$ is the quark-gluon vertex, the factor $(-1)$ signifies that the extra minus sign is due to the fermionic character of the quark loop, and the trace over color indices has already been performed.

The most general expression for the full quark propagator can be written as
\be\label{qprop}
S_F(p) = \frac{i Z(p^2)} { p \cdot \gamma - M(p^2)} \,,
\ee
where we shall refer to $Z(p^2)$ as the quark form factor, and $M(p^2)$ is the mass function.

Because of gauge invariance and multiplicative renormalizability we know that
\be\label{renormq}
g(\mu^2) = \frac{Z_3^{1/2}(\mu^2,\Lambda^2)
Z_2(\mu^2,\Lambda^2)}{Z_{1f}(\mu^2,\Lambda^2)} \, g_0(\Lambda^2)  \,\,, \hspace{1cm}
Z(p^2, \Lambda^2) = Z_2(\mu^2,\Lambda^2) Z_R(p^2, \mu^2) \,,
\ee
where $Z_{1f}$, $Z_2$ are the renormalization constants for the quark-gluon vertex and the quark field, and $Z_R$ is the renormalized quark form factor, with $Z_R(\mu^2, \mu^2) \equiv 1$.

The contribution of the quark-loop to Eq.~(\ref{SD5}) for the renormalized gluon form factor is computed from  Eq.~(\ref{piq}),  after contraction, Wick rotation, and multiplication with $Z_3$. It can always be written as
\be\label{qloop}
\pi^{q}_R(x) =  - N_f \frac{g^2(\mu^2)}{8 \pi^3} Z_{1f}^2(\mu^2,\Lambda^2)
\inttwo \,   V(x,y,z)\, Z_R(y) Z_R(z)  \,,
\ee
 where we introduced renormalized quantities using Eq.~(\ref{renormq}), and 
\be\label{Vkernel}
V(p^2,q^2,r^2) = - \frac{1}{6 p^2 \,q^2\, r^2} \,  {\cal P}^{\mu\nu}(p)
\text{Tr} \l[ \Gamma^{qg,0}_{\mu}(-r,q,-p) \, q\cdot\gamma \, \Gamma^{qg}_{\nu}(q,-r,p) \, r\cdot\gamma \r] \,,\\
\ee
in the massless case. All the information about the full quark-gluon vertex is contained in the kernel $V$. Note that the minus sign in front of $\pi_{\mu\nu}^{q}$ in Eq.~(\ref{SD-Gl-full}) has been absorbed in the definition of $\pi^{q}_R$. Similarly to the treatment of the gluon loop, we eliminate the renormalization constant $Z_{1f}$ using Eqs.~(\ref{renorm}, \ref{renormg}, \ref{renormq}),  expressing the equality of the renormalized quark-gluon and ghost-gluon couplings, and find:
\be
\label{qloop2}
\pi^{q}_R(x) = - \frac{\alpha_\mu}{2 \pi^2} 
\inttwo \,  \left[ N_f \,\frac{Z(y,\Lambda^2)Z(z,\Lambda^2)}{G(y,\Lambda^2)G(z,\Lambda^2)} \, V(x,y,z) \right] \, G_R(y) G_R(z)  \,.
\ee
Hence, Eq.~(\ref{SD7}) remains valid, provided we replace the definition (\ref{Rxyz}) for the kernel $R$ by
\begin{equation}\label{Rxyzq}
R(x,y,z) = M(x,y,z) + \frac{F(y,\Lambda^2)F(z,\Lambda^2)}{G(y,\Lambda^2)G(z,\Lambda^2)} \, Q(x,y,z) +    
\frac{N_f}{N_c} \,\frac{Z(y,\Lambda^2)Z(z,\Lambda^2)}{G(y,\Lambda^2)G(z,\Lambda^2)} \, V(x,y,z)\,.
\end{equation}

As before the $\mu$ dependence enters only through the renormalized coupling and the renormalized gluon and ghost form factors, and the $\Lambda$ dependence of $V$ is such that the integrals are finite. We herein study the case of $N_f$ massless quarks, thereby avoiding the complications introduced by the presence of explicit and dynamically generated mass terms. Again we introduce the bare approximation $R_0$ to Eq.~(\ref{Rxyzq}), defined by Eq.~(\ref{bare-elems}) supplemented with
\be
\Gamma_{\mu}^{qg} = \Gamma_{\mu}^{qg,0} \,, \qquad \qquad  Z = Z_0 = 1 \,,
\ee
 such that $R_0 = M_0 + Q_0 + V_0$, with the quark loop component $V_0$ given in Eq.~(\ref{irquark}). 

The ultraviolet analysis can be done completely analogously to that presented for the quarkless case in Appendix~\ref{App:bare}, and it again leads to the high momentum behavior shown in Eq.~(\ref{uv}), but now with $\beta_0 = (11N_c - 2 N_f)/3$, $\gamma=(-13N_c/2+2 N_f)/3\beta_0$, and $\delta=-3N_c/4\beta_0$, thus correctly reproducing the leading order perturbative results. 

From the structure of Eq.~(\ref{qloop2}) it is clear that massless quark loops will also contribute to the leading infrared power in the gluon equation in our bare approximation, in the assumption of infrared power behaved gluon and ghost form factors.  
The consistency condition (\ref{consist}) is now modified as the gluon equation gets an extra term from Eq.~(\ref{qloop2}), proportional to $N_f$ times the quark loop contribution $\chi_{gl}^q(\kappa)$ given in Eq.~(\ref{nuglnf}):
\be\label{nugltot}
N_c \, \chi_{gh}(\kappa) = N_c \, \chi_{gl}(\kappa) + N_f \, \chi_{gl}^q(\kappa) \,.
\ee

\begin{figure}
\begin{center}
\includegraphics[width=10cm]{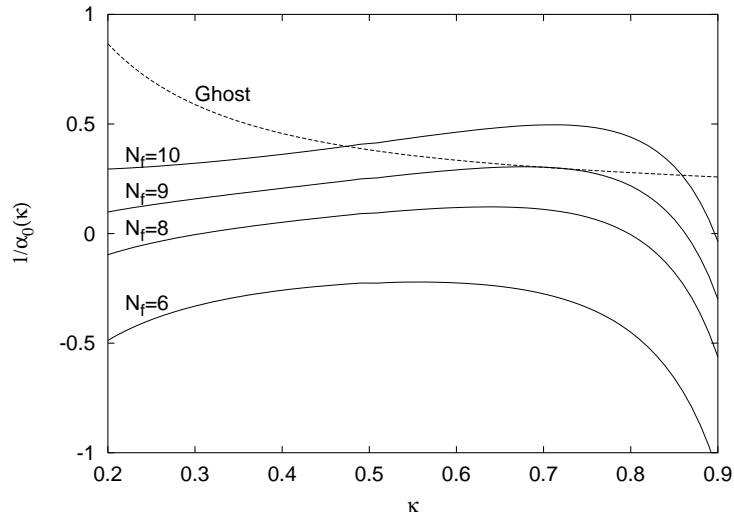}
\caption{\label{Fig-glnf}
Determination of the infrared exponent $\kappa$ as the intersection of the left and right hand sides of the consistency condition, Eq.~(\ref{nugltot}), computed from the ghost equation (dashed line) and gluon equation for $N_f=6,8,9,10$ (full lines).}
\end{center}
\end{figure}

We analyze the effect of the quark loop contribution on the right hand side of Eq.~(\ref{nugltot}), and observe that its contribution is positive. In Fig.~(\ref{Fig-glnf}), we show how it increases as we increase the number of massless quarks, and how it eventually compensates the negative value from the gluon loop (with $N_c=3$) such that Eq.~(\ref{nugltot}) be satisfied. There is a critical value $N_f \approx 8.9$ for which  a consistent infrared behavior can first be found: $\kappa \approx 0.76$. Then, when we increase $N_f$ further, two possible values of $\kappa$ satisfy the consistency condition, Eq.~(\ref{nugltot}). For $N_f=9$ we find $\kappa=0.69$ and $0.82$, for $N_f=10$: $\kappa=0.49$ and $0.90$, for $N_f=11$:  $\kappa=0.37$ and $0.93$, for $N_f=12$: $\kappa=0.28$ and $0.94$. As we keep increasing $N_f \to \infty$, the two solutions for $\kappa$ go to $0$ and $1$, the corresponding values for $\alpha_0$ are then $0$ and $4\pi/3$.

We conclude that, in the bare approximation, massless quarks do contribute to the leading infrared power of the gluon equation, and that a consistent power behavior can be found if the number of quarks is sufficiently large ($N_f \ge 9$), thus potentially restoring the existence of infrared power solutions for the gluon and ghost form factors. Of course, we know that the real world QCD does not contain such a large number of massless quarks, and the results of the infrared analysis will be altered, even in the bare approximation, by the presence of explicit quark masses. Further study is needed to investigate how massive quarks, with either explicit or dynamically generated quark masses, will influence these results. This will be done in future work in conjunction with a simultaneous solution of the quark Dyson-Schwinger equation. Note however, that the above-mentioned results might be of importance in the study of Yang-Mills Grand Unified Theories, which could contain a large number of massless quarks. 

More importantly, the previous calculation enabled us to see that, even though the multiplicatively renormalizable treatment of the gluon loop seemed to disprove the existence of infrared power law solutions, these can be reinstated by additional contributions to the vacuum polarization. 
As shown in the gluon equation, Eq.~(\ref{SD-Gl-full}), even the pure gauge theory, without quarks, contains additional diagrams. The three-gluon and four-gluon loop contributions have been neglected until now, based merely on perturbative arguments. We now schematically show that this is not justified, and that these diagrams can also contribute to the leading infrared behavior of the gluon and ghost propagators. 

The three-gluon contribution to the unrenormalized gluon equation, Eq.~(\ref{SD-Gl-full}), is
\be\label{pi3g}
\pi^{3g}_{\mu\nu}(p) =  \frac{N_c^2}{6}g_0^4 \int \frac{d^4 q}{(2\pi)^4} \int \frac{d^4 k}{(2\pi)^4} \, \Gamma^{4g,0}_{\mu\alpha_1\alpha_2\alpha_3}(-p,q,k,r) D^{\alpha_1\beta_1}(q) D^{\alpha_2\beta_2}(k) D^{\alpha_3\beta_3}(r) \Gamma^{4g}_{\nu\beta_3\beta_2\beta_1}(p,-r,-k,-q) \,,
\ee
where $r=p-q-k$.
After contraction, Wick rotation, and multiplication with $Z_3$ of Eq.~(\ref{pi3g}), the additional three-gluon contribution to Eq.~(\ref{SD7}) for the renormalized gluon form factor can be written in the form:
\be
\pi^{3g}_R(x) = -  Z_3 N_c^2 g_0^4 \int \frac{d^4 q}{(2\pi)^4}\int \frac{d^4 k}{(2\pi)^4} \, Q_3(p,q,k) F(z_1) F(z_2)  F(z_3) \,,
\ee
where $z_1, z_2, z_3$ are the squared momenta of the gluons in the loop. Introducing the renormalized coupling using Eq.~(\ref{renormg}), we can rewrite this as
\be
\pi^{3g}_R(x) = - N_c^2 g^4(\mu^2) \frac{\tilde Z_1^4}{Z_3 \tilde Z_3^4} \int \frac{d^4 q}{(2\pi)^4} \int \frac{d^4 k}{(2\pi)^4} \, Q_3(p,q,k)  F(z_1) F(z_2) F(z_3) \,.
\ee
As before, we replace all the renormalization constants by the ratios of the corresponding unrenormalized and renormalized form factors, and find
\be
\pi^{3g}_R(x) = -  16\pi^2 N_c^2 \alpha^2(\mu^2)\int \frac{d^4 q}{(2\pi)^4} \int \frac{d^4 k}{(2\pi)^4} \l[ \frac{  F(z_2)F(z_3)}{G(z_2) G(z_3) } \frac{Q_3(p,q,k) }{G^2(z_1)} \r]\, F_R(z_1) G_R^2(z_1) G_R(z_2) G_R(z_3) \,.
\ee

Using Eq.~(\ref{rgi2}) we recognize the running coupling at momentum $z_1$, $\alpha(z_1) = \alpha_\mu F_R(z_1) G_R^2(z_1)$, and can therefore rewrite this as
\be\label{3gloop}
\pi^{3g}_R(x) = - 16\pi^2 N_c^2 \alpha_\mu \int \frac{d^4 q}{(2\pi)^4} \int \frac{d^4 k}{(2\pi)^4}  \l[ \frac{F(z_2) F(z_3)}{G(z_2) G(z_3)} \frac{Q_3(p,q,k)}{ G^2(z_1)}\r] \,  \alpha(z_1) \, G_R(z_2) G_R(z_3) \,.
\ee
One of the effects of our manipulations has been to change one of the constant renormalized couplings $\alpha_\mu$ in front of the integral into a momentum dependent running coupling $\alpha(z_1)$ inside the integral.
In the assumption of infrared power laws, Eq.~(\ref{powlaw}), the running coupling has an infrared fixed point, and hence, the infrared structure of the integrand in Eq.~(\ref{3gloop}) is very similar to that of Eq.~(\ref{SD7}), and $\pi^{3g}_R(x)$ will potentially contribute to the leading infrared power analysis of the  gluon equation. 

We follow the same steps in the treatment of the four-gluon loop. Its contribution to Eq.~(\ref{SD-Gl-full}) is:
\ba
\pi^{4g}_{\mu\nu}(p) &=&  \frac{N_c^2}{4}\, g_0^4 \int \frac{d^4 q}{(2\pi)^4} \int \frac{d^4 k}{(2\pi)^4} \\ && \times \, \Gamma^{4g,0}_{\mu\alpha_1\alpha_2\alpha_3}(-p,q,k,r) D^{\alpha_1\beta_1}(q) D^{\alpha_2\beta_2}(k) D^{\alpha_3\beta_3}(r) \Gamma^{3g}_{\beta_1\alpha_4\beta_2}(-q,s,-k) D^{\alpha_4\beta_4}(s)  \Gamma^{3g}_{\nu\beta_3\beta_4}(p,-r,-s) \,, \nn
\ea
where $r=p-q-k, s= q+k$. After contraction, Wick rotation and multiplication with $Z_3$, the four-gluon contribution to Eq.~(\ref{SD7}) for the renormalized gluon form factor can be written as
\be
\pi^{4g}_R(x) = - Z_3 N_c^2 g_0^4  \int \frac{d^4 q}{(2\pi)^4}\int \frac{d^4 k}{(2\pi)^4} \, Q_4(p,q,k)  F(z_1) F(z_2) F(z_3) F(z_4) \,.
\ee
In analogy with Eq.~(\ref{3gloop}), we find
\be
\label{4gloop}
\pi^{4g}_R(x) = - 16\pi^2 N_c^2 \alpha_\mu  \int \frac{d^4 q}{(2\pi)^4} \int \frac{d^4 k}{(2\pi)^4}  \l[\frac{ F(z_2) F(z_3)}{ G(z_2) G(z_3)} \frac{Q_4(p,q,k) F(z_4)}{ G^2(z_1)}\r] \,  \alpha(z_1) G_R(z_2) G_R(z_3) \,,
\ee
and this diagram too can potentially contribute to the leading order infrared power. 

An infrared power law analysis of the gluon equation including the three-gluon and four-gluon contributions, Eqs.~(\ref{3gloop}, \ref{4gloop}), will yield a condition
\be\label{condit2}
(N_c \alpha_0) \chi_{gl}(\kappa) + (N_c \alpha_0)^2 \chi_{gl}^{3g,4g}(\kappa) = 1 \,,
\ee
for the pure gauge theory,  where $\chi_{gl}^{3g,4g}$ is to be calculated from Eqs.~(\ref{3gloop}, \ref{4gloop}). This condition replaces the second relation in Eq.~(\ref{condit}), and is now quadratic in $N_c \alpha_0$. The solution of Eq.~(\ref{condit2}) for $N_c \alpha_0$, together with the first part of Eq.~(\ref{condit}), will yield a new consistency equation
\be\label{consist2}
\chi_{gh}(\kappa) = \frac{1}{2}\left( \chi_{gl}(\kappa) + \sqrt{\chi_{gl}^2(\kappa) + 4 \chi_{gl}^{3g,4g}(\kappa)} \right)\,,
\ee
replacing Eq.~(\ref{consist}). The contributions of Eqs.~(\ref{3gloop}, \ref{4gloop}) involve the computation of two-loop integrals, and the approximations to the kernels have to be devised and investigated carefully. Note that these kernels too are independent of $\mu$ and are such that all integrals in the subtracted equations are finite.  It is crucial for the survival of the power law solutions as infrared asymptotic behavior of the ghost and gluon propagators, that a leading infrared exponent $\kappa$ is found which satisfies the condition, Eq.~(\ref{consist2}). 
A more detailed  analysis of Eqs.~(\ref{3gloop}, \ref{4gloop}) will be performed in future work.

Finally, we briefly turn our attention to the quark Dyson-Schwinger equation, in the massless case, and examine how it transforms in our new approach. The equation is:
\be\label{quark-dse}
[S_F(p)]^{-1} = [S_F^0(p)]^{-1}  
- C_F g_0^2 \int \frac{d^4 q}{(2\pi)^4} \, \Gamma^{qg,0}_\mu(p,q,r) S_F(q) \Gamma^{qg}_\nu(q,p,-r) D^{\mu\nu}(r)  \,,
\ee
where $C_F=(N_c^2-1)/2N_c=4/3$, for $N_c=3$.

After substituting Eq.~(\ref{qprop}) in Eq.~(\ref{quark-dse}), we can derive the following formula for the quark form factor:
\be\label{Zx}
\frac{1}{Z(x)} = 1 - \frac{C_F g_0^2}{8\pi^3} \inttwo \, U(x,y,z) Z(y) F(z) \,,
\ee
where the kernel $U$ depends on the full quark-gluon vertex.
Multiplying Eq.~(\ref{Zx}) by $Z_2$, and using Eq.~(\ref{renormg}) to introduce the renormalized coupling, we obtain
\be
\frac{1}{Z_R(x)} = Z_2(\mu^2,\Lambda^2) - \frac{C_F g^2(\mu^2)}{8\pi^3} \frac{ \tilde Z_1^2 Z_2}{ Z_3\tilde Z_3^2} \inttwo \, U(x,y,z) \, Z(y) F(z) \,.
\ee
As before, we remove all the renormalization constants in the self-energy term:
\be
\frac{1}{Z_R(x)} = Z_2(\mu^2,\Lambda^2) - \frac{C_F \alpha_\mu}{2\pi^2} \inttwo \, \l[\frac{Z^2(y)}{G^2(z)} \, U(x,y,z) \r]\,  F_R(z) G_R^2(z) \frac{1}{Z_R(y)} \,.
\ee
Similarly to the derivation of Eq.~(\ref{3gloop}), we recognize the running coupling $\alpha(z)$ and find:
\be\label{Zx2}
\frac{1}{Z_R(x)} = Z_2(\mu^2, \Lambda^2) + \frac{C_F}{2\pi^2} \inttwo \, \l[\frac{Z^2(y)}{G^2(z)} \, U(x,y,z)\r] \,  \alpha(z) \, \frac{1}{Z_R(y)} \,.
\ee
Subtraction of this equation at two momenta allows us to eliminate the remaining renormalization constant $Z_2$. Note that the multiplicative renormalizability of the solutions is manifested in a very simple way by the $1/Z_R$ structure on both sides of the equation. In fact, the shape of the equation is somewhat reminiscent of the massless, quenched QED, fermion equation derived by Curtis and Pennington\cite{Curtis:1991fb} using a vertex extension\cite{Curtis:1990zs} to the Ball-Chiu vertex\cite{Ball:1980ay}, constructed using conditions of multiplicative renormalizability.  Although we leave a detailed study of the quark equation to future work, we already see that we can introduce a bare approximation to  Eq.~(\ref{Zx2}), in analogy to the procedure applied to the gluon equation, which will yield solutions  respecting multiplicative renormalizability.

Furthermore, we anticipate that, in a similar way, the scalar part $M(p^2)$ of the quark propagator will also be driven by the running coupling under the integral, and it will be interesting to study how dynamical chiral symmetry breaking gets realized with a kernel proportional with $\alpha(z)$, which has an infrared fixed point if the infrared power laws, Eq.~(\ref{powlaw}), are verified. This will also be investigated in more detail in a further publication. 

Note that we have used the running coupling $\alpha(x) = \alpha_\mu F_R(x,\mu^2) G_R^2(x,\mu^2)$ in the derivations of Eqs.~(\ref{3gloop}, \ref{4gloop}, \ref{Zx2}), but we could as well introduce the more general renormalization group invariant quantity $\alpha(x,y,z) = \alpha_\mu F_R(x,\mu^2) G_R(y,\mu^2) G_R(z,\mu^2)$.
Although this will change the appearance of the equations, it will not change their solutions, as long as no approximations to the kernels are introduced. However, when devising truncations to the kernels, the choice of momenta should be handled judiciously for the approximations to make sense.

\section*{Conclusions}

We have reformulated the coupled set of continuum equations for the renormalized gluon and ghost form factors in QCD, such that the multiplicative renormalizability of the solutions is manifest, independently of the specific form of full vertices and renormalization constants. In the Landau gauge, all the renormalization constants are eliminated, and the renormalization point dependence is completely transposed into the renormalized coupling and the renormalized form factors. 

Furthermore, the specific structure of the equations allows us to devise novel approximations, necessary to make them tractable, with solutions which respect the principles of multiplicative renormalizability and have the correct leading order perturbative limit. This had not been achieved in a consistent way before.

We showed that the gluon and ghost equations are each individually satisfied by infrared power behaved gluon and ghost propagators. However, these infrared power solutions can only truly be identified with the QCD propagators if both equations are satisfied simultaneously, thereby determining the value of the leading infrared exponent and of the infrared fixed point of the coupling. 

In contrast with previous research, our study shows that the leading infrared behavior of the gluon equation is {\it not} solely determined by the ghost loop. The new approach shows that the contributions of the gluon loop, the three-gluon loop, the four-gluon loop and even of massless quarks should be taken into account when performing the leading order infrared analysis. 

 Moreover, we showed that, in our new Landau gauge truncation, the gluon loop contribution to the gluon vacuum polarization removes the existence of a consistent infrared power solution. It is our hope that additional contributions to the gluon vacuum polarization will reinstate the power law solutions. As an illustration we showed how including a large number of massless quarks is one way to recover consistency. The three-gluon and four-gluon loops will also contribute to the leading infrared power behavior in this approximation, but its quantitative treatment is more complicated and requires further investigation. We are therefore not yet able to conclude whether or not the gluon and ghost propagators are power behaved in the infrared.

We have also briefly shown, for the massless case, how the quark equation can be treated in a similar way, and how the specific structure of the equation will allow us to construct truncations with multiplicatively renormalizable solutions for the quark propagator.

\begin{acknowledgments}
 I thank  D.~Atkinson, C.~Fisher, K.~Langfeld, C.~D.~Roberts, S.~M.~Schmidt, and P.~Watson for useful comments and discussions. I am grateful for the hospitality of Argonne National Laboratory, where part of this work was completed. This work was funded by Deutsche Forschungsgemeinschaft under project no. SCHM 1342/3-1. 
\end{acknowledgments}

\appendix

\section{Bare approximation}
\label{App:bare}

 The bare approximation to Eqs.~(\ref{SD9}, \ref{SD10}), defined by Eqs.~(\ref{bare-elems}, \ref{bare-approx}), yields the following kernels:
\begin{eqnarray}\label{Rxyz0}
R_0(x,y,z) &=& - \frac{1}{3} \left[ {\frac{{x^2}}{8\,y^2\,z^2}}  
 +  {\frac{x}{y\,z}} \left( {\frac{{1}}{y}} + {\frac{1}{z}}\right)
 - {\frac{1}{8}} \left( {\frac{15}{y^2}}  + {\frac{34}{y\,z}}   +  {\frac{15}{z^2}} \right) \right.
\\ &&   + {\frac{1}{4\,x}}\left(   {\frac{z}{y^2}} - {\frac{11}{y}}  - {\frac{11}{z}}  + {\frac{{y}}{z^2}} 
\right)
 + \left. {\frac{1}{2\,x^2}} \left( 
  {\frac{{z^2}}{y^2}} + {\frac{6\,z}{y}} - 14  + {\frac{6\,{y}}{z}} + {\frac{{y^2}}{z^2}}  \right) \right] \nn \,, \\
\label{Txyz0}
T_0(x,y,z) &=&  - \left({\frac{{x}}{y}} -
      2 + {\frac{{y}}{x}}\right){\frac{1}{4\,z^2}} + 
  \left({\frac{1}{y}} + {\frac{1}{x}}\right){\frac{1}{2\,z}} - {\frac{1}{4\,x\,y}} = \frac{\sin^2\theta}{z^2} \,.
\end{eqnarray}

\subsection*{Ultraviolet behavior}

We follow the arguments of Ref.~\cite[Section (VII)]{Atkinson:1998tu} to study the ultraviolet behavior of Eqs.~(\ref{SD9}, \ref{SD10}) with kernels, Eqs.~(\ref{Rxyz0}, \ref{Txyz0}). 
The leading ultraviolet order solutions satisfy
\ba
\frac{1}{F_R(x)} &=& 1 - \frac{\alpha_\mu}{4\pi} \int_x^{\mu^2} dy \, \frac{f_0}{y} \, G_R^2(y) \,,\\
\frac{1}{G_R(x)} &=& 1 - \frac{\alpha_\mu}{4\pi} \int_x^{\mu^2} dy \, \frac{g_0}{y} \, F_R(y) G_R(y) \,,
\ea
where $f_0$ and $g_0$ are easily computed from the kernels. We have chosen the renormalization scale $\mu$ as subtraction point, and both $x$ and $\mu$ are in the perturbative region. It is easy to see that this is solved by:
\be\label{pertsol}
F_R(x) = \left( 1 + \frac{\beta_0\alpha_\mu}{4\pi}\log\frac{x}{\mu^2} \right)^{\gamma} \,,\qquad \qquad
G_R(x) = \left( 1 + \frac{\beta_0\alpha_\mu}{4\pi}\log\frac{x}{\mu^2} \right)^{\delta} \,,
\ee
where $\beta_0 = f_0+2 g_0$, $\gamma=-f_0/\beta_0$, and $\delta=-g_0/\beta_0$. This leads to a running coupling:
\be
\alpha(x) = \alpha_\mu F_R(x) G_R^2(x) = \frac{\alpha_\mu}{ 1 + \frac{\beta_0\alpha_\mu}{4\pi}\log\frac{x}{\mu^2} } = \frac{4\pi}{\beta_0\log\frac{x}{\Lambda_{QCD}^2}} \,,
\ee
where we define 
\be
\Lambda^2_{QCD} \equiv \mu^2 \exp\left( -\frac{4\pi}{\beta_0\alpha_\mu} \right)
\ee
to leading order. Hence, the solutions (\ref{pertsol}) can be rewritten as
\be
F_R(x,\mu^2) = \left( \frac{\alpha(x)}{\alpha_\mu} \right)^{-\gamma} \sim \log^{\gamma} x \,,\qquad \qquad
G_R(x,\mu^2) = \left( \frac{\alpha(x)}{\alpha_\mu} \right)^{-\delta} \sim \log^{\delta} x \,,
\ee
which is in agreement with the renormalization group equation results. In the bare approximation we find $\beta_0 = 11N_c/3$,  $\gamma=-13N_c/6\beta_0=-13/22$, and $\delta=-3N_c/4\beta_0=-9/44$.

\subsection*{Infrared behavior}

 In this section we give the results for $\chi_{gh}(\kappa)$ and $\chi_{gl}(\kappa)$ of Eq.~(\ref{condit}).
They are determined from the coefficients of the leading power of $x$ for $x \to 0$, after substitution of the power laws (\ref{powlaw}) in Eqs.~(\ref{SD9}, \ref{SD10}), and solution of the integrals.

 The angular and radial integrals are readily solved  using the method of Ref.~\cite{Atkinson:1998zc}, and the $x$-dependence of a typical integral is given by:
\be\label{integral}
\int_0^{\Lambda^2} \!\!\! dy \, y^b  \int_0^\pi \!\!\! d\theta \, \sin^{2j} z^a \sim H_j(a,b) \, x^{a+b} \,,
\ee
with $z = x+y-2\sqrt{x y} \cos\theta$, and
\ba\label{Hjab}
H_j(a,b) \equiv B\left(j+\frac{1}{2},\frac{1}{2}\right) & \times & \bigg[ \frac{1}{b+1} \, \FThreeTwo(-a,-a-j,b+1; j+1,b+2; 1) 
\\  && - \frac{1}{a+b+1}  \, \FThreeTwo(-a,-a-j,-a-b-1; j+1,-a-b; 1) \bigg] \,, \nn
\ea
where $B$ is the Beta-function, $\FThreeTwo$ is a generalized hypergeometric function.
The integrations are performed using the integration formula (3.665.2) of Ref.~\cite{Grads} and the definition of hypergeometric functions\cite{Abram}.

Using Eq.~(\ref{integral}) it is straightforward to see that the ghost equation, Eq.~(\ref{SD10}), with bare kernel (\ref{Txyz0}), yields:
\be
 \label{nugh}
\chi_{gh}(\kappa) = 
{-\frac{1}{2\pi^2}}\, H_2(2\,\kappa - 2 ,1 - \kappa ) \,.
\ee

The gluon equation, Eq.~(\ref{SD9}), with kernel (\ref{Rxyz0}), gives a similar expression: 
\ba
\label{nugl}
\chi_{gl}(\kappa) &=& {\frac{1}{48\pi^2}} \Big[{H_1}(-2 - \kappa ,-1 - \kappa ) - 
     15\,{H_1}(-2 - \kappa ,1 - \kappa ) + 
     2\,{H_1}(-2 - \kappa ,2 - \kappa ) \\ && + 
     4\,{H_1}(-2 - \kappa ,3 - \kappa ) + 
     8\,{H_1}(-2 - \kappa ,-\kappa ) + 
     8\,{H_1}(-1 - \kappa ,-1 - \kappa ) - 
     22\,{H_1}(-1 - \kappa ,1 - \kappa ) \nn\\ && + 
     24\,{H_1}(-1 - \kappa ,2 - \kappa ) - 
     34\,{H_1}(-1 - \kappa ,-\kappa ) + 
     2\,{H_1}(1 - \kappa ,-1 - \kappa ) + 
     24\,{H_1}(1 - \kappa ,-\kappa ) \nn\\ && + 
     4\,{H_1}(2 - \kappa ,-1 - \kappa ) - 
     15\,{H_1}(-\kappa ,-1 - \kappa ) - 
     56\,{H_1}(-\kappa ,1 - \kappa ) - 
     22\,{H_1}(-\kappa ,-\kappa )\Big] \nn \,.
\ea

\subsection*{Including massless quarks}

The bare, massless quark approximation $V_0$, to the kernel $V$ of Eq.~(\ref{qloop2}), is given by:
\be\label{irquark}
V_0(x,y,z) = -\frac{2}{3} \left[ {\frac{1}{y\,z}} + {\frac{1}{x}} \left( {\frac{1}{y}}+{\frac{1}{z}}\right) 
 -  {\frac{2}{x^2}} \left(  {\frac{y}{z}} - 2  + {\frac{z}{{y}}} \right) \right] \,.
\ee

After substituting the power solutions, Eq.~(\ref{powlaw}), in the quark loop, Eq.~(\ref{qloop2}), and solving the angular and radial integrals, we find a quark contribution:
\ba
\label{nuglnf}
\chi_{gl}^q(\kappa) &=& {\frac{1}{3\pi^2}} \,\Big[ {H_1}(-1 - \kappa ,1 - \kappa ) - 
       2\,{H_1}(-1 - \kappa ,2 - \kappa ) + {H_1}(-1 - \kappa ,-\kappa ) - 
       2\,{H_1}(1 - \kappa ,-\kappa ) \\ && + 4\,{H_1}(-\kappa ,1 - \kappa ) + 
       {H_1}(-\kappa ,-\kappa ) \Big] \nn
\ea
per massless quark flavor. 

\bibliography{glgh3}

\end{document}